\documentclass[a4paper,12pt,eqno]{article}
\usepackage{theorem}
\usepackage{latexsym,amssymb,amsfonts,amsmath}
\usepackage{graphicx}
\newtheorem{thm}{Theorem}[section]
\newtheorem{lem}[thm]{Lemma}
\newtheorem{cor}[thm]{Corollary}
\newtheorem{pro}[thm]{Proposition}
\newtheorem{rem}[thm]{Remark}

\newtheorem{exm}[thm]{Example}

\theoremheaderfont{\scshape}
\newcommand{\RM}{\mathbb{R}}

\newcommand{\CM}{\mathbb{C}}

\newcommand{\HM}{\mathbb{H}}
\newcommand{\Mat}{\operatorname{Mat}}
\newcommand{\RP}{\operatorname{Re}}
\newcommand{\CP}{\operatorname{Co}}
\newcommand{\IP}{\operatorname{Im}}

\title{{\Large {\bf The discrete-time quaternionic quantum walk \\on a graph}}
\author{
{\small Norio Konno}\\
{\scriptsize Department of Applied Mathematics, 
Faculty of Engineering, 
Yokohama National University}\\
{\scriptsize Hodogaya, Yokohama 240-8501, Japan}\\
{\scriptsize e-mail: konno@ynu.ac.jp
}\\
{\small Hideo Mitsuhashi}\\
{\scriptsize Faculty of Education, 
Utsunomiya University}\\
{\scriptsize Utsunomiya, Tochigi 321-8505, Japan}\\
{\scriptsize e-mail: mitsu@cc.utsunomiya-u.ac.jp
}\\
{\small Iwao Sato}\\
{\scriptsize Oyama National College of Technology}\\
{\scriptsize Oyama, Tochigi 323-0806, Japan}\\
{\scriptsize e-mail: isato@oyama-ct.ac.jp
}\\}
}
\date{\empty }
\begin{document}
\maketitle

\par\noindent
\begin{small}
\par\noindent
{\bf Abstract}. Recently, the quaternionic quantum walk was formulated by the first author as a generalization of discrete-time quantum walks. 
We treat the right eigenvalue problem of quaternionic matrices to analysis the spectra of its transition matrix. 
The way to obtain all the right eigenvalues of a quaternionic matrix is given. 
From the unitary condition on the transition matrix of the quaternionic quantum walk, 
we deduce some properties about it. 
Our main results, Theorem \ref{EigenFormula}, determine all the right eigenvalues of a quaternionic quantum walk by use of those of 
the corresponding weighted matrix. 
In addition, we give some examples of quaternionic quantum walks and their right eigenvalues. 
  
\footnote[0]{
{\it Abbr. title:} The discrete-time quaternionic quantum walk on a graph
}
\end{small}

\setcounter{equation}{0}
\section{Introduction}
As a quantum version of the classical random walk, the quantum walk has recently been studied in various fields. The discrete-time quantum walk 
in one dimension lattice was intensively studied by Ambainis et al. \cite{AmbainisEtAl2001}. One of the most striking properties is the spreading property of the walker. The standard deviation of the position grows linearly in time, quadratically faster than classical random walk. On the other hand, a walker stays at the starting position, i.e., localization occurs. 
Konno \cite{Konno2015} extended the quantum walk to a {\it quaternionic quatum walk} determined by a unitary matrix whose component is quaternion, and presented some properties of the walk. The reviews and books on quantum walks are Kempe \cite{Kempe2003}, Konno \cite{Konno2008b}, Venegas-Andraca \cite{Venegas2013}, Cantero et al. \cite{CGMV}, Manouchehri and Wang \cite{MW2013}, Portugal \cite{Port2013} for examples. 

For a general graph, a typical case of discrete-time quantum walks is the Grover walk on a graph. 
The Grover walk on a graph was formulated in \cite{Grover1996}.
Emms et al. \cite{EmmsETAL2006} treated spectra of the transition matrix and its positive support of 
the Grover walk on a graph, and showed that the third power of the transition matrix 
outperforms the graph spectra methods in distinguishing strongly regular graphs. 
Godsil and Guo \cite{GG2010} gave new proofs of the results of Emms et al. \cite{EmmsETAL2006}.  

The discrete-time quantum walk on a graph is closely related to the Ihara
zeta function of a graph. 
Zeta functions of graphs originally started from the Ihara zeta function for a regular graph by Ihara \cite{Ihara1966}. 
The Ihara zeta function of a graph was studied by many researchers \cite{Ihara1966}, \cite{Serre}, 
\cite{Sunada1986}, \cite{Sunada1988}, \cite{Hashimoto1989}, \cite{Bass1992}, \cite{ST1996}, \cite{FZ1999}, 
\cite{KS2000}. 
Bass \cite{Bass1992} generalized Ihara's result on the zeta function of 
a regular graph to an irregular graph, and showed that its reciprocal is 
again a polynomial.
Inspired by these works, Sato \cite{Sato} defined a new zeta function (the second weighted zeta function) of a graph 
by using not an infinite product but a determinant. This new zeta function and its 
determinant formula played an essential role in the determination of eigenvalues of quantum walk on a graph 
in Konno and Sato \cite{KS2012}. 
Ren et al.  \cite{RenETAL} found an interesting relation between
the Ihara zeta function and the Grover walk on a graph, and
showed that the positive support of the transition matrix of the Grover walk is equal to 
the Perron-Frobenius operator (the edge matrix) related to the Ihara zeta function. 
Konno and Sato \cite{KS2012} gave the characteristic polynomials of the transition matrix of the Grover walk and its
positive support by using the second weighted zeta function and the Ihara zeta function, and so 
obtained the other proofs of the results on spectra for them by Emms et al. \cite{EmmsETAL2006}.

Quaternion, which can be considered as an extension of complex number, was
discovered by Hamilton in 1843. 
The eigenvalue problem for matrices whose entries are quaternions has been investigated 
for nearly a century by a number of researchers. 
A detailed survey on the quaternionic matrices can be found in Zhang \cite{Zhang1997}. 
The notable difference from the eigenvalue problem for 
complex matrices is that it is necessary to treat left eigenvalues and right eigenvalues 
separately. Right eigenvalues are well studied by, for example, Brenner \cite{Brenner1951} 
and Lee \cite{Lee1949}. 
On the contrary, left eigenvalues are less known and not easy to handle as commented in 
Huang and So \cite{HS2001}. 
Recently, an extension of quantum walk to the case of quaternions was established by Konno \cite{Konno2015} in which a model of quaternionic quantum walk  was suggested and its some properties were presented. In the present paper, we extend a discrete-time quantum walk (Grover walk) on a graph to a walk given by a 
unitary matrix whose components are quaternions (quaternionic Grover walk). 
We derive the condition for components of transition matrix of quaternionic Grover walk. 
Furthermore, we deal with right eigenvalue problem, and derive all the right eigenvalues of 
quaternionic transition matrix from those of the corresponding weighted matrix. 
Our results can be regarded as a generalization of \cite{EmmsETAL2006}, \cite{KS2012}.

The rest of the paper is organized as follows. 
Section 2 treats the right eigenvalue problem of quaternionic matrices. 
The way to obtain all the right eigenvalues of a quaternionic matrix is given. 
In Section 3, we provide a summary of the Ihara zeta function and the second weighted zeta function of a graph, and present their determinant expressions. 
In Section 4, we show a brief account of the discrete-time quantum walk on a graph, and define a generalization of it to the case of quaternions, which we call 
the quaternionic quantum walk on a graph. From the unitary condition on the transition matrix, 
we deduce some properties about it. 
In Section 5, 
our main results, Theorem \ref{EigenFormula}, determine all the right eigenvalues of a quaternionic quantum walk by use of those of the corresponding weighted matrix. 
In addition, we give some examples of quaternionic quantum walks and their right eigenvalues.

\section{Right eigenvalues of a quaternionic matrix}
In this section, we shall show the way to calculate all the right eigenvalues 
of a quaternionic matrix. Since quaternions do not mutually commute in general, 
we must treat left eigenvalue and right eigenvalue separately. 
Since ${\bf M}^n{\bf v}={\bf v}{\lambda}^n\;
(n{\;\geq\;}1)$ holds for an right eigenvalue $\lambda$ of {\bf M} and an eigenvector 
{\bf v} corresponding to $\lambda$, right eigenvalue is more suitable to calculate 
the stationary measure of the quaternionic quantum walk than left eigenvalue. 
Therefore, we treat only right eigenvalues in this paper. 
Although we cannot apply the ordinary characteristic polynomial for a 
quaternionic matrix, we can calculate all the complex right eigenvalues of it 
by embedding it in a complex matrix twice the size of it. 
From these eigenvalues, we can obtain all the right eigenvalues of the 
quaternionic matrix. 
An exposition of these contents can be found in \cite{Aslaksen1996}. 
\cite{Zhang1997} gives an overview of quaternionic matrix theory.  

Let $\HM$ be the set of quaternions. $\HM$ is a noncommutative associative 
algebra over $\RM$, whose underlying real vector space has dimension $4$ 
with a basis $1,i,j,k$ which satisfy the following relations: 
\[
i^2=j^2=k^2=-1,\quad ij=-ji=k,\quad jk=-kj=i,\quad ki=-ik=j.
\]
For $x=x_0+x_1i+x_2j+x_3k{\;\in\;}\HM$, we call $x_0$ the {\it real part}, 
$x_0+x_1i$ the {\it complex part}, $x_1i+x_2j+x_3k$ the {\it imaginary part }of $x$, and 
set $\RP\,x=x_0$, $\CP\,x=x_0+x_1i$, $\IP\,x=x_1i+x_2j+x_3k$ respectively. 
$x^*$ denotes the {\it conjugate} of $x$ in $\HM$ 
which is defined by $x^*=x_0-x_1i-x_2j-x_3k$.
We call $|x|=\sqrt{xx^*}=\sqrt{x^*x}=\sqrt{x_0^2+x_1^2+x_2^2+x_3^2}$ the {\it norm} of $x$. 
Indeed, $|{\,\cdot\,}|$ satisfies 

(1) $|x|{\;\geq\;}0$, and moreover $|x|=0 \Leftrightarrow x=0$,

(2) $|xy|=|x||y|$,

(3) $|x+y|{\;\leq\;}|x|+|y|$.\\
For a nonzero element $x{\;\in\;}\HM$, $x^{-1}=x^*/|x|^2$. 
Hence, $\HM$ constitutes a skew field. 

We can present arbitrary quaternion $x$ by two complex numbers 
$x=a+jb$ uniquely. Such a presentation is called 
{\it symplectic decomposition}. 
Two complex numbers $a$ and $b$ are called {\it simplex part} and 
{\it perplex part} 
of $x$ respectively. 
Symplectic decomposition is also valid for a quaternionic matrix, namely 
a matrix whose entries are quaternions. 
$\Mat(m{\times}n,\HM)$ denotes the set of $m{\times}n$ quaternionic matrices and 
$\Mat(n,\HM)$ the set of $n{\times}n$ quaternionic square matrices. 
For ${\bf M}{\;\in\;}\Mat(m{\times}n,\HM)$, we can write ${\bf M}={\bf A}+j{\bf B}$ 
uniquely where ${\bf A},{\bf B}{\;\in\;}\Mat(m{\times}n,\CM)$. 
${\bf A}$ and ${\bf B}$ are called {\it simplex part} and 
{\it perplex part} 
of ${\bf M}$ respectively. 
We define $\psi$ to be the map from $\Mat(m{\times}n,\mathbb{H})$ to 
$\Mat(2m{\times}2n,\mathbb{C})$ as follows: 
\[
\psi : \Mat(m{\times}n,\mathbb{H}){\;\longrightarrow\;}\Mat(2m{\times}2n,\mathbb{C})
\quad{\bf M}{\;\mapsto\;}\begin{bmatrix}{\bf A}&-\overline{\bf B}\\{\bf B}&
\overline{\bf A}\end{bmatrix},
\]
where $\overline{\bf A}$ is the complex conjugate of ${\bf A}$. 
Then $\psi$ is an $\RM$-linear map. We also have: 

\begin{lem}\label{PsiLem}
Let ${\bf M}{\;\in\;}\Mat(m{\times}n,\HM)$ and ${\bf N}{\;\in\;}
\Mat(n{\times}m,\HM)$. Then 
\[ \psi({\bf M}{\bf N})=\psi({\bf M})\psi({\bf N}). \]
\end{lem}

{\bf Proof}. 
Let ${\bf M}={\bf A}+j{\bf B}$ and ${\bf N}={\bf C}+j{\bf D}$ 
be symplectic decompositions of ${\bf M}$ and ${\bf N}$. Then, 
\[ {\bf M}{\bf N}=({\bf A}+j{\bf B})({\bf C}+j{\bf D})
={\bf A}{\bf C}+{\bf A}j{\bf D}+j{\bf B}{\bf C}+j{\bf B}j{\bf D}. \]
Since  
${\bf X}j=j\overline{\bf X}$ for arbitrary complex matrix ${\bf X}$, 
it follows that: 
\[ {\bf M}{\bf N}={\bf A}{\bf C}-\overline{{\bf B}}{\bf D}
+j(\overline{{\bf A}}{\bf D}+{\bf B}{\bf C}), \]
and therefore
\[ \psi({\bf M}{\bf N})
=\begin{bmatrix}{\bf A}{\bf C}-\overline{{\bf B}}{\bf D}
&-{\bf A}\overline{{\bf D}}-\overline{{\bf B}}\overline{{\bf C}}\\
\overline{{\bf A}}{\bf D}+{\bf B}{\bf C}
&\overline{{\bf A}}\overline{{\bf C}}-{\bf B}\overline{{\bf D}}
\end{bmatrix}.
\]
On the other hand, 
\[ \psi({\bf M})\psi({\bf N})=
\begin{bmatrix}{\bf A}&-\overline{{\bf B}}\\
{\bf B}&\overline{{\bf A}}
\end{bmatrix}
\begin{bmatrix}{\bf C}&-\overline{{\bf D}}\\
{\bf D}&\overline{{\bf C}}
\end{bmatrix}
=\begin{bmatrix}{\bf A}{\bf C}-\overline{{\bf B}}{\bf D}
&-{\bf A}\overline{{\bf D}}-\overline{{\bf B}}\overline{{\bf C}}\\
{\bf B}{\bf C}+\overline{{\bf A}}{\bf D}
&-{\bf B}\overline{{\bf D}}+\overline{{\bf A}}\overline{{\bf C}}
\end{bmatrix}.
\]
Thus $\psi({\bf M}{\bf N})=\psi({\bf M})\psi({\bf N})$ holds. 
$\Box$

\begin{pro}
If $m=n$, then $\psi$ is an injective $\mathbb{R}$-algebra homomorphism. 
\end{pro}

{\bf Proof}. 
By Lemma \ref{PsiLem}, $\psi$ is an $\RM$-algebra homomorphism. 
Injectivity of $\psi$ is clear. 
$\Box$

For a quaternionic square matrix ${\bf M}=({\bf M}_{ij}){\;\in\;}\Mat(n,\HM)$, 
the {\it conjugate} ${\bf M}^*=(({\bf M}^*)_{ij})$ is defined by 
$({\bf M}^*)_{ij}=({\bf M}_{ji})^*$. We notice that 
$\psi({\bf M}^*)=\psi({\bf M})^*$, where the right-hand side denotes the conjugate 
transpose of the complex matrix. 
A quaternionic square matrix ${\bf M}$ is said to be {\it unitary} if 
${\bf M}^*{\bf M}={\bf M}{\bf M}^*={\bf I}$. Since $\psi$ is an injective 
$\RM$-algebra homomorphism, ${\bf M}^*{\bf M}=I$ implies ${\bf M}{\bf M}^*=I$.

\begin{rem}
$\psi({}^T\!{\bf M})={}^T\!\psi({\bf M})$ does not hold in general, 
where ${}^T\!{\bf M}$ denotes the transpose of ${\bf M}$. 
\end{rem}

Let $\mathbb{H}^n$ be a right vector space, and 
${\bf M}{\;\in\;}\Mat(n,\mathbb{H}),\;{\bf v}{\;\in\;}\mathbb{H}^n,\;\lambda{\;\in\;}\mathbb{H}$ 
satisfy ${\bf M}{\bf v}={\bf v}{\lambda}$. 
$\lambda$ is said to be a right eigenvalue of ${\bf M}$, and ${\bf v}$ 
a right eigenvector corresponding to $\lambda$. 
We denote by $\sigma_r({\bf M})$ the set of all the right eigenvalues of ${\bf M}$. 
We have ${\bf M}({\bf v}q)={\bf v}{\lambda}q={\bf v}q(q^{-1}{\lambda}q)$ 
for any $q{\;\in\;}\mathbb{H}^*=\mathbb{H}-\{0\}$. Hence, the conjugate class 
$\lambda^{\mathbb{H}^*}=\{q^{-1}{\lambda}q\;|\;q{\;\in\;}\mathbb{H}^*\}$ is contained in $\sigma_r({\bf M})$. 
We notice that if $\lambda{\;\in\;}\mathbb{R}$, then $\lambda^{\mathbb{H}^*}=\{\lambda\}$. 
Since $|q^{-1}xq|=|x|$, the group homomorphism: 
\[
\rho : \mathbb{H}^*{\;\longrightarrow\;}\operatorname{Aut}\mathbb{H},\qquad
\rho(q)x=qxq^{-1}
\]
gives orthogonal transformations $\rho(q)\;(q{\;\in\;}\mathbb{H}^*)$ on 
$\mathbb{H}$. Let $\mathbb{H}_R=\mathbb{R}$ and $\mathbb{H}_P=\mathbb{R}i{\oplus}
\mathbb{R}j{\oplus}\mathbb{R}k$. 
Then $\mathbb{H}=\mathbb{H}_R{\oplus}\mathbb{H}_P$ and 
$\mathbb{H}_R{\;\perp\;}\mathbb{H}_P$. 
We call the elements of $\mathbb{H}_P$ {\it pure} quaternion. 
One can see that $\rho(q)x=x\;(x{\;\in\;}\mathbb{H}_R)$ and 
$\rho(q)\mathbb{H}_P=\mathbb{H}_P$, hence 
$\rho$ also gives orthogonal transformations $\rho(q)\;(q{\;\in\;}\mathbb{H}^*)$ on 
$\mathbb{H}_P$. Moreover it is well-known that $\det(\rho(q))=1$ 
and that the group homomorphism:
\[
\rho|_{\mathbb{H}_P} : \mathbb{H}^*{\;\longrightarrow\;}SO(\mathbb{H}_P)
\cong SO(3)
\]
is surjective (for details, see for example \cite{Altmann2005},\cite{CS2003}). 
Therefore, $\mathbb{H}^*$ gives all the 
rotations in $\mathbb{H}_P{\;\cong\;}\mathbb{R}^3$. 

If $\lambda{\;\in\;}\mathbb{H}-\mathbb{R}$, then we may write 
$\lambda=\lambda_0+\lambda'$ ($\lambda_0{\;\in\;}\mathbb{H}_R,\;\lambda'{\;\in\;}
\mathbb{H}_P$) with $\lambda'{\;\neq\;}0$. Then 
$\rho(q){\lambda}=q{\lambda}q^{-1}=\lambda_0+q{\lambda'}q^{-1}$. 
It follows that $\lambda'^{\mathbb{H}^*}$ is a $2$-sphere centering at the origin 
in $\mathbb{H}_P$ since the following equations hold: 

\begin{equation}\label{QuatRot}
\lambda'^{\mathbb{H}^*}=\{q^{-1}{\lambda'}q\;|\;q{\;\in\;}\mathbb{H}^*\}
=\rho(\mathbb{H}^*){\lambda'}
=SO(\mathbb{H}_P){\lambda'}. 
\end{equation}

Thus the intersection of $\lambda'^{\mathbb{H}^*}$ and the complex axis consists of 
just two points, $\lambda'^{\mathbb{H}^*}{\cap}\mathbb{R}i=\{{\pm}|\lambda'|i\}$. 
To summarize our discussion, we find out the following:  
\begin{pro}\label{QuatConj}
Let ${\bf M}{\;\in\;}\Mat(n,\HM)$. 
The ${\mathbb{H}^*}$-conjugate class of any right eigenvalue of ${\bf M}$ 
which is not real is contained in $\sigma_r({\bf M})$, and 
has exactly two complex numbers that are conjugate.
\end{pro} 

We shall show that one can calculate all the complex right eigenvalues 
of a quaternionic matrix by use of ordinary determinant. 

\begin{pro}\label{EigenCorres}
Let ${\bf M}{\;\in\;}\Mat(n,\HM)$ and $p{\;\in\;}\mathbb{C}$. Then 
\begin{equation*}
{\bf M}{\bf v}={\bf v}p{\;\Leftrightarrow\;}\psi({\bf M})
\begin{bmatrix} {\bf u} \\ {\bf w} \end{bmatrix}
=\begin{bmatrix} {\bf u} \\ {\bf w} \end{bmatrix}p
\qquad \text{where\quad} {\bf v}={\bf u}+j{\bf w}\;
({\bf u},{\bf w}{\;\in\;}\mathbb{C}^n).
\end{equation*}
\end{pro}

{\bf Proof}. 
Let ${\bf M}={\bf A}+j{\bf B}$ be the symplectic decomposition. Then we have: 
\[ {\bf M}{\bf v}=({\bf A}+j{\bf B})({\bf u}+j{\bf w})
=({\bf A}{\bf u}-\overline{\bf B}{\bf w})
+j({\bf B}{\bf u}+\overline{\bf A}{\bf w}). \]
Hence ${\bf M}{\bf v}={\bf v}p$ is equivalent to the following equation: 
\[ 
\begin{bmatrix}{\bf A}&-\overline{\bf B}\\{\bf B}&\overline{\bf A}\end{bmatrix}
\begin{bmatrix}{\bf u} \\ {\bf w}\end{bmatrix}
=\begin{bmatrix}{\bf u} \\ {\bf w}\end{bmatrix}p.
\]
$\Box$

In the same way as the ordinary triangularization of a complex matrix,  
the triangularization of a quaternionic matrix can also be obtained. 
Moreover, we can triangularize 
a quaternionic matrix into a triangular matrix whose diagonals are complex 
as follows. The detail of the proof can be found in \cite{Lee1949}. 

\begin{lem}\label{ComplexEigen}
For arbitrary ${\bf M}{\;\in\;}\Mat(n,\HM)$, there exist 
${\bf V}{\;\in\;}\Mat(n,\HM)$ with ${\bf V}^*{\bf V}={\bf V}{\bf V}^*={\bf I}_n$ 
and $\lambda_1,{\ldots},\lambda_n{\;\in\;}\CM$ such that 
\begin{equation}\label{Triangularization}
{\bf V}^*{\bf M}{\bf V}=
\begin{bmatrix}
\lambda_1 & * &{\cdots}& * \\
0 & \lambda_2 & & * \\
{\vdots}& & \ddots &{\vdots}\\
0 & 0 &{\cdots}&\lambda_n
\end{bmatrix}.
\end{equation}
\end{lem}

As a consequence of Proposition \ref{EigenCorres} and Lemma 
\ref{ComplexEigen}, we can deduce the following: 

\begin{pro}\label{EigenPair}
Let ${\bf M}{\;\in\;}\Mat(n,\HM)$. 
Then $2n$ eigenvalues of $\psi({\bf M}){\;\in\;}\Mat(2n,\CM)$ occur 
in conjugate pairs: 
\[
\lambda_1,\overline{\lambda_1},\lambda_2,\overline{\lambda_2},
{\ldots},\lambda_n,\overline{\lambda_n},
\]
counted with multiplicity. 
\end{pro}

{\bf Proof}. 
The proof can be found in \cite{Lee1949}. 
Let ${\bf T}$ be the right hand side of (\ref{Triangularization}) and 
${\bf T}={\bf T}^S+j{\bf T}^P$ its symplectic decomposition. 
Then, 
\begin{equation*}
\begin{split}
\det(\lambda{\bf I}_{2n}-\psi({\bf M}))
&=\det(\lambda{\bf I}_{2n}-\psi({\bf V})^{-1}\psi({\bf M})\psi({\bf V}))\\
&=\det(\lambda{\bf I}_{2n}-\psi({\bf V}^*{\bf M}{\bf V}))\\
&=\det(\lambda{\bf I}_{2n}-\psi({\bf T}))
=\det \begin{bmatrix}\lambda{\bf I}_{2n}-{\bf T}^S & \overline{{\bf T}^P}\\
-{\bf T}^P & \lambda{\bf I}_{2n}-\overline{{\bf T}^S}
\end{bmatrix}\\
&=\det \begin{bmatrix}
\lambda-\lambda_1& * & {\cdots} & * & 0 & * & {\cdots} & * \\
0 &\lambda-\lambda_2&  & * & 0 & 0 & & * \\
{\vdots} &   & \ddots & {\vdots} & {\vdots} &   & \ddots & {\vdots} \\
0 & 0 & {\cdots} &\lambda-\lambda_n & 0 & 0 & {\cdots} & 0 \\
0 & * & {\cdots} & * & \lambda-\overline{\lambda_1}& * & {\cdots} & * \\
0 & 0 &  & * & 0 &\lambda-\overline{\lambda_2}& & * \\
{\vdots} &   & \ddots & {\vdots} & {\vdots} &   & \ddots & {\vdots} \\
0 & 0 & {\cdots} & 0 & 0 & 0 & {\cdots} &\lambda-\overline{\lambda_n}
\end{bmatrix}\\
&=(\lambda-\lambda_1)(\lambda-\overline{\lambda_1}){\cdots}
(\lambda-\lambda_n)(\lambda-\overline{\lambda_n}).
\end{split}
\end{equation*}
$\Box$

Hence, diagonal elements in ${\bf T}$ and their complex conjugates 
are eigenvalues of $\psi({\bf M})$. 
We notice that we may replace $\lambda_i$ by its complex conjugate in 
(\ref{Triangularization}) because of Proposition \ref{QuatConj}. 
Combining Proposition \ref{QuatConj}, \ref{EigenCorres}, \ref{EigenPair}, 
we can state about eigenvalues of a quaternionic matrix as follows:

\begin{thm}\label{QuatEigen}
For arbitrary quaternionic matrix ${\bf M}{\;\in\;}\Mat(n,\mathbb{H})$, 
$2n$ complex right eigenvalues $\lambda_1,{\ldots},\lambda_n,$ 
$\overline{\lambda_1},{\ldots},\overline{\lambda_n}$ of ${\bf M}$ exist 
counted with multiplicity,
which can be calculated by solving $\det({\lambda}{\bf I}_{2n}-\psi({\bf M}))=0$. 
The set of right eigenvalues $\sigma_r({\bf M})$ is given by 
$\sigma_r({\bf M})=\lambda_1^{\mathbb{H}^*}{\cup}{\;\cdots\;}{\cup}\lambda_n^{\mathbb{H}^*}$. 
\end{thm}

\begin{exm}
${\bf M}=\begin{bmatrix}0&1\\-1&0\end{bmatrix}$\\
$\det({\lambda}{\bf I}_{4}-\psi({\bf M}))=\det\begin{bmatrix}\lambda&-1&0&0\\1&\lambda&0&0\\
0&0&\lambda&-1\\0&0&1&\lambda\end{bmatrix}=(\lambda-i)^2(\lambda+i)^2$. \quad
$\sigma_r({\bf M})=i^{\mathbb{H}^*}$.
\end{exm}

\begin{exm}
${\bf M}=\begin{bmatrix}1&0\\0&i\end{bmatrix}$\\
$\det({\lambda}{\bf I}_{4}-\psi({\bf M}))=\det\begin{bmatrix}\lambda-1&0&0&0\\0&\lambda-i&0&0\\
0&0&\lambda-1&0\\0&0&0&\lambda+i\end{bmatrix}=(\lambda-1)^2(\lambda-i)(\lambda+i)$. \\
$\sigma_r({\bf M})=\{1\}{\cup}i^{\mathbb{H}^*}$.
\end{exm}

\begin{exm}
${\bf M}=\begin{bmatrix}0&i\\j&0\end{bmatrix}=\begin{bmatrix}0&i\\0&0\end{bmatrix}
+j\begin{bmatrix}0&0\\1&0\end{bmatrix}$\\
$\det({\lambda}{\bf I}_{4}-\psi({\bf M}))=\det\begin{bmatrix}\lambda&-i&0&0\\0&\lambda&1&0\\
0&0&\lambda&i\\-1&0&0&\lambda\end{bmatrix}=(\lambda^2-i)(\lambda^2+i)$\\
$=\left( \lambda-\dfrac{1+i}{\sqrt{2}} \right)
\left(\lambda-\dfrac{-1-i}{\sqrt{2}} \right)\left(\lambda-\dfrac{1-i}{\sqrt{2}} \right)\left(\lambda-\dfrac{-1+i}{\sqrt{2}} \right)$. \\ 
$\sigma_r({\bf M})=\Big{(}\dfrac{1+i}{\sqrt{2}}\Big{)}^{\mathbb{H}^*}{\cup}
\Big{(}\dfrac{-1-i}{\sqrt{2}}\Big{)}^{\mathbb{H}^*}$.
\end{exm}

\section{The Ihara zeta function of a graph} 
Let $G=(V(G)$, $E(G))$ be a finite connected graph with the set $V(G)$ of 
vertices and the set $E(G)$ of unoriented edges $uv$ 
joining two vertices $u$ and $v$. 
We assume that $G$ has neither loops nor multiple edges throughout. 
For $uv \in E(G)$, an arc $(u,v)$ is the oriented edge from $u$ to $v$. 
Let $D(G)=\{\,(u,v),\,(v,u)\,\mid\,uv{\;\in\;}E(G)\}$ and 
$|V(G)|=n,\;|E(G)|=m,\;|D(G)|=2m$. 
For $e=(u,v){\;\in\;}D(G)$, $o(e)=u$ denotes the {\it origin} and $t(e)=v$ the {\it terminal} 
of $e$ respectively. 
Furthermore, let $e^{-1}=(v,u)$ be the {\em inverse} of $e=(u,v)$. 
The {\em degree} $\deg v = \deg {}_G \  v$ of a vertex $v$ of $G$ is the number of edges 
incident to $v$. 
A {\em path $P$ of length $\ell$} in $G$ is a sequence 
$P=(e_1, \cdots ,e_{\ell})$ of $\ell$ arcs such that $e_i \in D(G)$ and 
$t(e_i)=o(e_{i+1})$ for $i{\;\in\;}\{1,\cdots,\ell-1\}$. 
We set $o(P)=o(e_1)$ and $t(P)=t(e_{\ell})$. $|P|$ denotes the length of $P$. 
We say that a path $P=(e_1, \cdots ,e_{\ell})$ has a {\em backtracking} 
if $ e_{i+1} =e_i^{-1} $ for some $i(1{\leq}i{\leq}\ell-1)$, and that 
$P=(e_1, \cdots ,e_{\ell})$ has a {\em tail} if $ e_{\ell} =e_1^{-1} $. 
A path $P$ is said to be a {\em cycle} if $t(P)=o(P)$. 
The {\em inverse} of a path 
$P=(e_1,\cdots,e_{\ell})$ is the path 
$(e_{\ell}^{-1},\cdots,e_1^{-1})$ and is denoted by $P^{-1}$.

We introduce an equivalence relation between cycles. 
Two cycles $C_1 =(e_1, \cdots ,e_{\ell})$ and 
$C_2 =(f_1, \cdots ,f_{\ell})$ are said to be {\em equivalent} if there exists 
$k$ such that $f_j =e_{j+k}$ for all $j$ where indices are treated modulo $\ell$. 
Let $[C]$ be the equivalence class which contains the cycle $C$. 
Let $B^r$ be the cycle obtained by going $r$ times around a cycle $B$. 
Such a cycle is called a {\em power} of $B$. 
A cycle $C$ is said to be {\em reduced} if 
$C$ has no backtracking and no tail. 
Furthermore, a cycle $C$ is said to be {\em prime} if it is not a power of 
a strictly smaller cycle.

The {\em Ihara zeta function} of a graph $G$ is 
a function of $t \in {\bf C}$ with $|t|$ sufficiently small, 
defined by 
\[
{\bf Z} (G, t)= {\bf Z}_G (t)= \prod_{[C]} (1- t^{ \mid C \mid } )^{-1} ,
\]
where $[C]$ runs over all equivalence classes of prime, reduced cycles 
of $G$. 

Let $G$ be a connected graph with $n$ vertices and $m$ unoriented edges. 
Two $2m \times 2m$ matrices 
${\bf B}=({\bf B}_{ef})_{e,f \in D(G)} $ and 
${\bf J}_0=( {\bf J}_{ef} )_{e,f \in D(G)} $ 
are defined as follows: 
\[
{\bf B}_{ef} =\left\{
\begin{array}{ll}
1 & \mbox{if $t(e)=o(f)$, } \\
0 & \mbox{otherwise,}
\end{array}
\right.
{\bf J}_{ef} =\left\{
\begin{array}{ll}
1 & \mbox{if $f= e^{-1} $, } \\
0 & \mbox{otherwise.}
\end{array}
\right.
\]
Then the matrix ${\bf B} - {\bf J}_0 $ is called the {\em edge matrix} of $G$.

\begin{thm}[Hashimoto \cite{Hashimoto1989}; Bass \cite{Bass1992}]
Let $G$ be a connected graph. 
Then the reciprocal of the Ihara zeta function of $G$ is given by 
\[
{\bf Z} (G, t)^{-1} =\det ( {\bf I}_{2m} -t ( {\bf B} - {\bf J}_0 ))
=(1- t^2 )^{r-1} \det ( {\bf I}_n -t {\bf A}+ 
t^2 ({\bf D} -{\bf I}_n )), 
\]
where $r$ and ${\bf A}$ are the Betti number and the adjacency matrix 
of $G$, respectively, and ${\bf D} =({\bf D}_{uv})_{u,v{\in}V(G)}$ is the diagonal matrix 
with ${\bf D}_{uu} = \deg u $ for arbitrary $u{\;\in\;}V(G)$. 
\end{thm}

Consider an $n \times n$ complex matrix 
${\bf W} =({\bf W}_{uv})_{u,v{\in}V(G)}$ with $(u,v)$-entry 
equals $0$ if $(u,v){\;\notin\;}D(G)$. 
We call ${\bf W}$ an {\em weighted matrix} of $G$.
Furthermore, let $w(u,v)= {\bf W}_{uv}$ for $u,v \in V(G)$ and 
$w(e)= w(u,v)$ if $e=(u,v) \in D(G)$.

For an weighted matrix ${\bf W}$ of $G$, a $2m \times 2m$ complex matrix 
${\bf B}_w=( {\bf B}^{(w)}_{ef} )_{e,f \in D(G)}$ is defined as follows: 
\begin{equation*}
{\bf B}^{(w)}_{ef} =\left\{
\begin{array}{ll}
w(f) & \mbox{if $t(e)=o(f)$, } \\
0 & \mbox{otherwise.}
\end{array}
\right.
\end{equation*}
Then the {\em second weighted zeta function} of $G$ is defined by 
\[
{\bf Z}_1 (G,w,t)= \det ( {\bf I}_{2m} -t ( {\bf B}_w - {\bf J}_0 ) )^{-1} . 
\]
If $w(e)=1$ for any $e \in D(G)$, then the second weighted zeta function of $G$ 
coincides with the Ihara zeta function of $G$.

\begin{thm}[Sato \cite{Sato}]\label{SatoThm}
Let $G$ be a connected graph, and 
let ${\bf W}$ be a weighted matrix of $G$. 
Then the reciprocal of the second weighted zeta function of $G$ is given by 
\[
{\bf Z}_1 (G,w,t )^{-1} =(1- t^2 )^{m-n} 
\det ({\bf I}_n -t {\bf W}+ t^2 ( {\bf D}_w - {\bf I}_n )) , 
\]
where $n=|V(G)|$, $m=|E(G)|$ and 
${\bf D}_w =({\bf D}^{(w)}_{uv})_{u,v{\in}V(G)}$ is the diagonal matrix 
with ${\bf D}^{(w)}_{uu} = \displaystyle \sum_{e:o(e)=u}w(e)$ for arbitrary 
$u{\;\in\;}V(G)$. 
\end{thm}

\begin{rem}
For later use, we mention that 
taking transpose, the following equation also holds: 
\[
\det ( {\bf I}_{2m} -t ( {}^T\!{\bf B}_w - {\bf J}_0 ) ) =(1- t^2 )^{m-n} 
\det ({\bf I}_n -t {}^T\!{\bf W}+ t^2 ( {\bf D}_w - {\bf I}_n )). 
\]
\end{rem}

\section{The transition matrix of a quaternionic quantum walk on a graph} 
A discrete-time quantum walk is a quantum process on a graph whose state vector is governed by 
a unitary matrix called the transition matrix. An important example of the quantum walk on a graph 
is the Grover walk, which was introduced in \cite{Grover1996}. 
Let $G$ be a connected graph with $n$ vertices and $m$ edges. 
Set $d_u = \deg u $ for $u{\;\in\;}V(G)$. 
The {\em transition matrix} of the Grover walk ${\bf U}^{\rm Gro}=
({\bf U}^{\rm Gro}_{ef})_{e,f \in D(G)} $ 
of $G$ is defined by 
\[
{\bf U}^{\rm Gro}_{ef} =\left\{
\begin{array}{ll}
2/d_{t(f)} (=2/d_{o(e)} ) & \mbox{if $t(f)=o(e)$ and $f \neq e^{-1} $, } \\
2/d_{t(f)} -1 & \mbox{if $f= e^{-1} $, } \\
0 & \mbox{otherwise.}
\end{array}
\right. 
\]
${\bf U}$ is called the {\it Grover matrix}. 
We denote by $Spec ({\bf A})$ the multiset of spectra of a complex square matrix 
${\bf A}$ counted with multiplicity. 
Let ${\bf T}=({\bf T}_{uv})_{u,v \in V(G)}$ be the $n \times n$ matrix defined as follows: 
\[
{\bf T}_{uv} =\left\{
\begin{array}{ll}
1/d_u  & \mbox{if $(u,v) \in D(G)$, } \\
0 & \mbox{otherwise.}
\end{array}
\right.
\] 
In \cite{EmmsETAL2006}, Emms et al. determined the spectra of ${\bf U}^{\rm Gro}$ 
by use of those of ${\bf T}$.

\begin{thm}[Emms, Hancock, Severini and Wilson \cite{EmmsETAL2006}] \label{EigenGro}
Let $G$ be a connected graph with $n$ vertices and $m$ edges. 
The transition matrix ${\bf U}^{\rm Gro}$ has $2n$ eigenvalues of the form: 
\[
\lambda = \lambda {}_{\bf T} \pm i \sqrt{1- \lambda {}^2_{\bf T} } , 
\]
where $\lambda {}_{\bf T} $ is an eigenvalue of the matrix ${\bf T}$. 
The remaining $2(m-n)$ eigenvalues of ${\bf U}^{\rm Gro}$ are $\pm 1$ 
with equal multiplicities. 
\end{thm}

Now, we extend the Grover walk to the case of quaternions. 
A discrete-time quaternionic quantum walk is a quantum process on a graph whose 
state vector, whose entries are quaternions, 
is governed by a quaternionic unitary matrix called the quaternionic transition matrix.  
Let $G$ be a finite connected graph with $n$ vertices and $m$ edges, 
The {\em quaternionic transition matrix} ${\bf U} =( {\bf U}_{ef} )_{e,f \in D(G)} $ 
of $G$ is defined by 
\begin{equation*}
{\bf U}_{ef} =\left\{
\begin{array}{ll}
q(e) & \mbox{if $t(f)=o(e)$ and $f \neq e^{-1} $, } \\
q(e) -1 & \mbox{if $f= e^{-1} $, } \\
0 & \mbox{otherwise,}
\end{array}
\right.
\end{equation*}
where $q$ is a map from $D(G)$ to $\HM$. 
The unitary condition on ${\bf U}$ is equivalent to the following equations: 

\begin{lem}
Let $q(e)=q_0(e)+q_1(e)i+q_2(e)j+q_3(e)k$. Then ${\bf U}$ is unitary 
if and only if the following five equations hold:
\begin{eqnarray}
&q_0(e)^2+q_1(e)^2+q_2(e)^2+q_3(e)^2-\dfrac{2q_0(e)}{d_{o(e)}}=0,\label{EqnUni1}\\
&(q_0(e)q_0(f)+q_1(e)q_1(f)+q_2(e)q_2(f)+q_3(e)q_3(f))d_{o(e)}-(q_0(e)+q_0(f))=0,
\label{EqnUni2-1}\\
&(-q_0(e)q_1(f)+q_1(e)q_0(f)-q_2(e)q_3(f)+q_3(e)q_2(f))d_{o(e)}-(q_1(e)-q_1(f))=0,
\label{EqnUni2-2}\\
&(-q_0(e)q_2(f)+q_2(e)q_0(f)-q_3(e)q_1(f)+q_1(e)q_3(f))d_{o(e)}-(q_2(e)-q_2(f))=0,
\label{EqnUni2-3}\\
&(-q_0(e)q_3(f)+q_3(e)q_0(f)-q_1(e)q_2(f)+q_2(e)q_1(f))d_{o(e)}-(q_3(e)-q_3(f))=0,
\label{EqnUni2-4}
\end{eqnarray}
where $e$ and $f$ are different arcs and have the same origin. 
\end{lem}

{\bf Proof}.  Observing that 
\[
({\bf U}{\bf U}^*)_{ef}=\sum_{g{\in}D(G),t(g)=o(e)=o(f)}{\bf U}_{eg}({\bf U}_{fg})^*,
\]
one can readily check by direct calculations that 
$({\bf U}{\bf U}^*)_{ee}=1$ is equivalent to (\ref{EqnUni1}), and 
$({\bf U}{\bf U}^*)_{ef}=0$ to (\ref{EqnUni2-1}),(\ref{EqnUni2-2}),(\ref{EqnUni2-3}) 
and (\ref{EqnUni2-4}). 
If $o(e){\;\neq\;}o(f)$, then clearly $({\bf U}{\bf U}^*)_{ef}=0$ holds. 
$\Box$

From (\ref{EqnUni1}), it follows that $q_0(e)$ must satisfy 
\begin{equation*}
0{\;\leq\;}q_0(e){\;\leq\;}\dfrac{2}{d_{o(e)}}. 
\end{equation*}

\begin{thm}\label{EqualityOfWeights}
If ${\bf U}$ is unitary, then $q(e)=q(f)$ holds for any two arcs $e,f{\;\in\;}D(G)$ 
with $o(e)=o(f)$. 
\end{thm}

{\bf Proof}.  We set $q_0(e)=x/d_u,\,q_0(f)=y/d_u$ ($0{\;\leq\;}x,y{\;\leq\;}2$) 
and ${\bf q}(e)={}^T\!(q_1(e),q_2(e),q_3(e))$, 
where $u=o(e)=o(f)$. 
Then (\ref{EqnUni2-1}) turns into:
\begin{equation}\label{EqnUni2'-1}
\dfrac{(x-1)(y-1)-1}{d_u}+d_u{\bf q}(e){\cdot}{\bf q}(f)=0
\end{equation}
On the other hand, (\ref{EqnUni2-2}),(\ref{EqnUni2-3}),(\ref{EqnUni2-4}) can be expressed 
all at once as follows: 
\begin{equation}\label{EqnUni2'-234}
(y-1){\bf q}(e)-(x-1){\bf q}(f)-d_u{\bf q}(e){\times}{\bf q}(f)=0.
\end{equation}
We show the assertion divided into three cases depending upon ${\bf q}(e)$ and ${\bf q}(f)$. 

If ${\bf q}(e)={\bf q}(f)={\bf 0}$, then (\ref{EqnUni2'-234}) trivially holds and 
(\ref{EqnUni2'-1}) turns into $(x-1)(y-1)-1=0$. 
In addition, by (\ref{EqnUni1}) we have 
\[
\dfrac{x(x-2)}{d_u}=0
\]
Thus, $(x,y)$ must be $(0,0)$ or $(2,2)$. 

If only either one of ${\bf q}(e)$ or ${\bf q}(f)$ equals ${\bf 0}$, say ${\bf q}(f)={\bf 0}$, 
then (\ref{EqnUni2'-234}) becomes $(y-1){\bf q}(e)={\bf 0}$. Hence $y=1$ and 
$q(f)=1/d_u$. But from (\ref{EqnUni2'-1}) it follows that $-1/d_u=0$, thereby 
reaching a contradiction. 

If both ${\bf q}(e)$ and ${\bf q}(f)$ do not equal ${\bf 0}$, 
then by (\ref{EqnUni2'-234}) it follows that:
\begin{eqnarray}
&(y-1){\bf q}(e)-(x-1){\bf q}(f)={\bf 0}\label{EqnBothNot0-1},\\
&{\bf q}(e){\times}{\bf q}(f)={\bf 0}\label{EqnBothNot0-2}.
\end{eqnarray}
By (\ref{EqnBothNot0-2}), ${\bf q}(e)$ and ${\bf q}(f)$ are parallel, so we may write 
${\bf q}(f)=c{\bf q}(e)$ for some nonzero real number $c$. 
Then (\ref{EqnBothNot0-1}) implies $y-1=c(x-1)$, and 
(\ref{EqnUni2'-1}) yields:
\begin{equation}\label{EqnOnC}
c\{(x-1)^2+d_u{}^2|{\bf q}(e)|^2\}=1
\end{equation}
Finally, substituting $x=q_0(e)d_u$ for (\ref{EqnOnC}) and using 
(\ref{EqnUni1}), we get $c=1$. 
$\Box$

Conversely, if $q(e)=q(f)$ for any two arcs $e,f{\;\in\;}D(G)$ with $o(e)=o(f)$, 
then (\ref{EqnUni2-1}) turns into (\ref{EqnUni1}), and moreover, (\ref{EqnUni2-2}), 
(\ref{EqnUni2-3}), (\ref{EqnUni2-4}) hold trivially. 
Thus we obtain: 

\begin{cor}
${\bf U}$ is unitary 
if and only if 
\begin{eqnarray*}
&q_0(e)^2+q_1(e)^2+q_2(e)^2+q_3(e)^2-\dfrac{2q_0(e)}{d_{o(e)}}=0,
\end{eqnarray*}
and $q(e)=q(f)$ for any two arcs $e,f{\;\in\;}D(G)$ with $o(e)=o(f)$. 
\end{cor}

Hereafter we impose the following condition on $q(e)$:   
\begin{equation}\label{QuatCond}
\displaystyle \sum_{e:o(e)=u}q(e) \text{ does not depend on $u$.} 
\end{equation}
Hence, we may write $\alpha=\alpha_0+\alpha_1i+\alpha_2j+\alpha_3k=\sum_{e:o(e)=u}q(e)$ 
for arbitrary $u{\;\in\;}V(G)$. 
Theorem \ref{EqualityOfWeights} and (\ref{QuatCond}) imply 
$q(e)=\alpha/d_{o(e)}$. 
If the imaginary part of $\alpha$ equals $0$, then ${\bf U}$ must equal either 
$-{\bf J}_0$ or ${\bf U}^{\rm Gro}$. 
Therefore, ${\bf U}$ can be regarded as a quaternionic extension of the Grover walk.

\section{Eigenvalues of the transition matrix of a 
quaternionic quantum walks on a graph}
We shall calculate all the right eigenvalues of ${\bf U}$. 
By Theorem \ref{QuatEigen}, it suffices to solve 
$\det({\lambda}{\bf I}_{4m}-\psi({\bf U}))=0$. 
Consider the set of quaternionic conjugates 
\[
{\bf U}^{\HM^*}=\{q^{-1}{\bf U}q\;|\;q{\;\in\;}\HM^*\},
\]
of ${\bf U}$. 
By (\ref{QuatRot}) and the subsequent argument, 
${\alpha}^{\HM^*}{\cap}\CM=\alpha_{\pm}$ where 
$\alpha_{\pm}=\alpha_0{\pm}\sqrt{\alpha_1^2+\alpha_2^2+\alpha_3^2}\,i$. 

Let ${\bf U}_{\pm}=(({\bf U}_{\pm})_{ef})_{e,f{\in}D(G)}$ be two complex matrices 
defined as follows: 
\begin{equation*}
({\bf U}_{\pm})_{ef} =\left\{
\begin{array}{ll}
\dfrac{\alpha_{\pm}}{d_{o(e)}} & \mbox{if $t(f)=o(e)$ and $f \neq e^{-1} $, } \\
\dfrac{\alpha_{\pm}}{d_{o(e)}} -1 & \mbox{if $f= e^{-1} $, } \\
0 & \mbox{otherwise.}
\end{array}
\right.
\end{equation*}
We notice that ${\bf U}_-=\overline{{\bf U}_+}$. 
Then, ${\bf U}^{\HM^*}{\cap}\Mat(2m,\CM)=\{{\bf U}_+,\,{\bf U}_-\}$. 
Therefore, for the quaternion $q$ such that $q^{-1}{\alpha}q=\alpha_+$, we have:
\begin{equation*}
\begin{split}
\det({\lambda}{\bf I}_{4m}-\psi({\bf U}))
&=\det(\psi(q{\bf I}_{2m})^{-1}({\lambda}{\bf I}_{4m}-\psi({\bf U}))\psi(q{\bf I}_{2m}))\\
&=\det({\lambda}{\bf I}_{4m}-\psi(q{\bf I}_{2m})^{-1}\psi({\bf U})\psi(q{\bf I}_{2m}))\\
&=\det({\lambda}{\bf I}_{4m}-\psi(q^{-1}{\bf U}q))\\
&=\det({\lambda}{\bf I}_{4m}-\psi({\bf U}_+)).
\end{split}
\end{equation*}
Thus, under the condition (\ref{QuatCond}), we can calculate all the right eigenvalues of 
${\bf U}$ by calculating eigenvalues of $\psi({\bf U}_+)$. However, 
\begin{equation}\label{DetFact}
\begin{split}
\det({\lambda}{\bf I}_{4m}-\psi({\bf U}_+))&=\det\Big{(}
\begin{bmatrix}\lambda{\bf I}_{2m}&{\bf O}\\{\bf O}&\lambda{\bf I}_{2m}\end{bmatrix}
-\begin{bmatrix}{\bf U}_+&{\bf O}\\{\bf O}&{\bf U}_-\end{bmatrix}\Big{)}\\
&=\det(\lambda{\bf I}_{2m}-{\bf U}_+)\det(\lambda{\bf I}_{2m}-{\bf U}_-).
\end{split}
\end{equation}
Hence, 
$Spec(\psi({\bf U}))=Spec(\psi({\bf U}_+))=
Spec({\bf U}_+){\cup}Spec({\bf U}_-)$. Since $Spec({\bf U}_-)$ is obtained 
from $Spec({\bf U}_+)$ by conjugating their elements, 
$Spec(\psi({\bf U}))$ 
consists of eigenvalues of ${\bf U}_+$ and their complex conjugates. 
Thus by Theorem \ref{QuatEigen}, we obtain:

\begin{thm}\label{EigenU}
Let ${\bf U}$ is a quaternionic transition matrix on a graph which satisfies 
(\ref{QuatCond}). Then $\sigma_r({\bf U})$ is given by: 
\[
\sigma_r({\bf U})=\lambda_1^{\HM^*}{\cup}{\cdots}{\cup}\lambda_n^{\HM^*},
\]
where $\lambda_1,{\cdots},\lambda_n$ are eigenvalues of ${\bf U}_+$. 
\end{thm}

Now we apply Theorem \ref{SatoThm} to the eigenvalue problem for ${\bf U}$. 
Let us define the $n{\times}n$ complex matrix 
${\bf W}_{\pm}=(({\bf W}_{\pm})_{uv})_{u,v{\in}V(G)}$ as follows: 
\begin{equation*}
({\bf W}_{\pm})_{uv} =\left\{
\begin{array}{ll}
\dfrac{\alpha_{\pm}}{d_u} & \mbox{if $(u,v){\;\in\;}D(G)$, } \\
0 & \mbox{otherwise.}
\end{array}
\right.
\end{equation*}

\begin{pro}\label{QuatDetFor}
Let $G$ be a connected graph with $n$ vertices and $m$ edges. 
Then, for the quaternionic transition matrix ${\bf U}$ of $G$,  
\begin{equation*}
\begin{split}
\det(\lambda{\bf I}_{2m}-{\bf U}_{\pm})
&=(\lambda{}^2-1)^{m-n}\det((\lambda^2+\alpha_{\pm}-1){\bf I}_{n}
-\lambda{}^T\!{\bf W}_{\pm}).
\end{split}
\end{equation*}
\end{pro}

{\bf Proof}.  Let $G$ be a connected graph with $n$ vertices and $m$ edges. 
Corresponding to ${\bf W}_{\pm}$, we define 
${\bf B}_{w{\pm}}=(({\bf B}^{(w)}_{\pm})_{ef})_{e,f{\in}D(G)}{\;\in\;}\Mat(2m,\CM)$ 
to be as follows: 
\[
({\bf B}^{(w)}_{\pm})_{ef} =\left\{
\begin{array}{ll}
\dfrac{\alpha_{\pm}}{d_{o(f)}} & \mbox{if $t(e)=o(f)$, } \\
0 & \mbox{otherwise.}
\end{array}
\right.
\]
We readily see that ${\bf U}_{\pm}={}^T\!{\bf B}_{w{\pm}}-{\bf J}_0$. 
Then, by Theorem \ref{SatoThm} and the subsequent remark, it follows that:  
\begin{equation*}
\begin{split}
\det ( {\bf I}_{2m} -t{\bf U}_{\pm})&=
\det ( {\bf I}_{2m} -t({}^T\!{\bf B}_{w{\pm}} - {\bf J}_0))\\
&=(1- t^2 )^{m-n} 
\det ({\bf I}_{n} -t{}^T\!{\bf W}_{\pm}+t^2({\bf D}_{w{\pm}} - {\bf I}_{n}))\\
&=(1- t^2 )^{m-n}\det ({\bf I}_{n} -t{}^T\!{\bf W}_{\pm}+ 
(\alpha_{\pm}-1)t^2 {\bf I}_{n} ) \\
&=(1- t^2 )^{m-n}\det ((1+(\alpha_{\pm}-1)t^2){\bf I}_{n} -t {}^T\!{\bf W}_{\pm}).
\end{split}
\end{equation*}
Now, let $t=1/ \lambda $. 
Then we have 
\[ 
\det \left( {\bf I}_{2m} - \frac{1}{ \lambda } {\bf U}_{\pm} \right)= 
\left(1- \frac{1}{ \lambda {}^2 } \right)^{m-n} 
\det \left(\left(1+(\alpha_{\pm}-1)\frac{1}{\lambda{}^2}\right){\bf I}_{n}
-\frac{1}{\lambda}{}^T\!{\bf W}_{\pm}\right). 
\]
This implies the following conclusion: 
\[
\det ( \lambda {\bf I}_{2m} - {\bf U}_{\pm} )= 
( \lambda {}^2 -1)^{m-n} \det (( \lambda {}^2 +\alpha_{\pm}- 1){\bf I}_{n} 
- \lambda {}^T\!{\bf W}_{\pm}) . 
\]
$\Box$

Using Proposition \ref{QuatDetFor}, we can calculate the spectra of $\psi({\bf U})$ 
as follows.

\begin{thm}\label{EigenFormula}
Let $G$ be a connected graph with $n$ vertices and $m$ edges. 
$Spec(\psi({\bf U}))$ has $4m$ elements. 
If $G$ is not a tree, then $4n$ eigenvalues of them are of the form: 
\begin{equation*}
\lambda = \dfrac{\mu \pm \sqrt{\mu^2-4(\alpha_+-1)}}{2},\; 
\dfrac{\nu \pm \sqrt{\nu^2-4(\alpha_--1)}}{2},
\end{equation*}
where $\mu{\;\in\;}Spec({}^T\!{\bf W}_+),\nu{\;\in\;}Spec({}^T\!{\bf W}_-)$. 
The remaining $4(m-n)$ eigenvalues of $\psi({\bf U})$ are $\pm 1$ 
with equal multiplicities. 
If $G$ is a tree, then $Spec(\psi({\bf U}))$ is as follows: 
\begin{equation*}
\begin{split}
&Spec(\psi ({\bf U}))\\
&=\Big{\{}\dfrac{\mu \pm \sqrt{\mu^2-4(\alpha_+-1)}}{2},\;
\dfrac{\nu \pm \sqrt{\nu^2-4(\alpha_--1)}}{2}\;
\Bigl|\; \mu {\in} Spec({}^T\!{\bf W}_+),\;
\nu{\in}Spec({}^T\!{\bf W}_-))\Big{\}}\\
&\ \ -\{1,1,-1,-1\}.
\end{split}
\end{equation*}
\end{thm}

{\bf Proof}.  By (\ref{DetFact}) and Proposition \ref{QuatDetFor}, we have 
\[
\det ( \lambda {\bf I}_{4m} - \psi({\bf U}) )= 
( \lambda {}^2 -1)^{2m-2n} 
\prod_{ \mu \in Spec({}^T\!{\bf W}_+)}(\lambda {}^2 +\alpha_+-1 -\mu \lambda)
\prod_{ \nu \in Spec({}^T\!{\bf W}_-)}(\lambda {}^2 +\alpha_--1 -\nu \lambda). 
\]
Solving $\lambda {}^2 +\alpha_+-1 -\mu \lambda =0$ and 
$\lambda {}^2 +\alpha_--1 -\nu \lambda =0$, we obtain 
\[
\lambda = \dfrac{\mu \pm \sqrt{\mu^2-4(\alpha_+-1)}}{2},\; 
\dfrac{\nu \pm \sqrt{\nu^2-4(\alpha_--1)}}{2}.
\]
If $G$ is not a tree, then $m{\;\geq\;}n$ and hence the statement clearly holds. 
If $G$ is a tree, then $m=n-1$. Since $\det ( \lambda {\bf I}_{4m} - \psi({\bf U}) )$ is 
a polynomial of $\lambda$, 
$\prod_{ \mu  \in Spec ({}^T\!{\bf W}_+)} ( \lambda {}^2 +\alpha_+-1 -\mu \lambda )\prod_{ \nu  \in Spec ({}^T\!{\bf W}_-)} ( \lambda {}^2 +\alpha_--1 -\nu \lambda )$ must 
have the factor $( \lambda {}^2 -1)^2$. Thus the statement holds for trees. 
$\Box$

\begin{rem}
Since $Spec({}^T\!{\bf W}_-)$ is obtained from $Spec({}^T\!{\bf W}_+)$ by 
conjugating their elements, it suffices to calculate only the eigenvalues of 
$Spec({}^T\!{\bf W}_+)$. 
\end{rem}

\begin{exm}
Let $G=K_3$, the complete graph with $3$ vertices, and 
$\alpha=1+1/2i+\sqrt{2}/2j-1/2k$. 
We give a weighted matrix as follows: 
\[
{\bf W}=\begin{bmatrix}
0&\frac{\alpha}{2}&\frac{\alpha}{2}\\
\frac{\alpha}{2}&0&\frac{\alpha}{2}\\
\frac{\alpha}{2}&\frac{\alpha}{2}&0
\end{bmatrix}.
\] 
Then the quaternionic transition matrix ${\bf U}$ is as follows: 
\[ 
{\bf U}={}^T\!{\bf B}_w-{\bf J}_0=\begin{bmatrix}
0&\frac{\alpha}{2}-1&0&0&\frac{\alpha}{2}&0\\
\frac{\alpha}{2}-1&0&0&\frac{\alpha}{2}&0&0\\
\frac{\alpha}{2}&0&0&\frac{\alpha}{2}-1&0&0\\
0&0&\frac{\alpha}{2}-1&0&0&\frac{\alpha}{2}\\
0&0&\frac{\alpha}{2}&0&0&\frac{\alpha}{2}-1\\
0&\frac{\alpha}{2}&0&0&\frac{\alpha}{2}-1&0
\end{bmatrix}.
\]
Then $\alpha_{\pm}=1{\pm}i$ and 
\[
Spec({}^T\!{\bf W}_+)
=\Bigg{\{}1+i,-\dfrac{1}{2}-\dfrac{1}{2}i,-\dfrac{1}{2}-\dfrac{1}{2}i\Bigg{\}},\;
Spec({}^T\!{\bf W}_-)
=\Bigg{\{}1-i,-\dfrac{1}{2}+\dfrac{1}{2}i,-\dfrac{1}{2}+\dfrac{1}{2}i\Bigg{\}}.
\]
and 
\begin{equation*}
\begin{split}
Spec({\bf U}_+)&=\Bigg{\{}1,i,
\dfrac{-1{\pm}\sqrt{7}}{4}+\dfrac{-1{\mp}\sqrt{7}}{4}i,
\dfrac{-1{\pm}\sqrt{7}}{4}+\dfrac{-1{\mp}\sqrt{7}}{4}i \Bigg{\}},\\
Spec({\bf U}_-)&=\Bigg{\{}1,-i,
\dfrac{-1{\pm}\sqrt{7}}{4}+\dfrac{1{\pm}\sqrt{7}}{4}i,
\dfrac{-1{\pm}\sqrt{7}}{4}+\dfrac{1{\pm}\sqrt{7}}{4}i \Bigg{\}}.
\end{split}
\end{equation*}
Thus we obtain: 
\begin{equation*}
\begin{split}
\sigma_r({\bf U})&=\{1\}
{\cup}i^{\HM^*}
{\cup}\Bigg{(}\dfrac{-1+\sqrt{7}}{4}+\dfrac{-1-\sqrt{7}}{4}i\Bigg{)}^{\!\!\HM^*}
{\cup}\Bigg{(}\dfrac{-1-\sqrt{7}}{4}+\dfrac{-1+\sqrt{7}}{4}i\Bigg{)}^{\!\!\HM^*}.
\end{split}
\end{equation*}

Next, we state spectra of the Grover matrix. 
Let $\alpha=2$. 
Then ${\bf U} $ is the Grover matrix ${\bf U}^{\rm Gro} $: 
\[ 
{\bf U}^{\rm Gro}=\begin{bmatrix}
0 & 0 & 0 & 0 & 1 & 0 \\
0 & 0 & 0 & 1 & 0 & 0 \\
1 & 0 & 0 & 0 & 0 & 0 \\
0 & 0 & 0 & 0 & 0& 1 \\
0 & 0 & 1 & 0 & 0 & 0 \\
0 & 1 & 0 & 0 & 0 & 0 
\end{bmatrix}.
\]
Furthermore, the transition matrix ${\bf T} $ of the simple random walk of $K_3 $ is 
given as follows: 
\[
{\bf T}=\begin{bmatrix}
0&\frac{1}{2}&\frac{1}{2}\\
\frac{1}{2}&0&\frac{1}{2}\\
\frac{1}{2}&\frac{1}{2}&0
\end{bmatrix}.
\] 
Then 
\[
Spec({\bf T})
=\Bigg{\{}1,-\dfrac{1}{2}, -\dfrac{1}{2}\Bigg{\}}
\quad \text{and} \quad  
Spec({\bf U}^{\rm Gro})=\Bigg{\{}1,1,
\dfrac{-1{\pm}\sqrt{3}i}{4}, \dfrac{-1{\pm}\sqrt{3}i}{4} \Bigg{\}}.
\]
Thus we obtain: 
\[
\sigma_r({\bf U}^{\rm Gro} )=\{1\} {\cup}  
\Bigg{(}\dfrac{-1+\sqrt{3}i}{2} \Bigg{)}^{\!\!\HM^*}.
\]
\end{exm}

\begin{exm}
Let $G=S_4$, the star graph with $4$ vertices, and 
$\alpha=4/3+1/3i+2/3j+\sqrt{3}/3k$. 
We give a weighted matrix as follows: 
\[
{\bf W}=\begin{bmatrix}
0&0&0& \alpha \\
0&0&0& \alpha \\
0&0&0&\alpha\\
\frac{\alpha}{3}&\frac{\alpha}{3}&\frac{\alpha}{3}&0
\end{bmatrix}.
\] 
Then the quaternionic transition matrix ${\bf U}$ is as follows: 
\[ 
{\bf U}={}^T\!{\bf B}_w-{\bf J}_0=\begin{bmatrix}
0&\alpha-1&0&0&0&0\\
\frac{\alpha}{3}-1&0&\frac{\alpha}{3}&0&\frac{\alpha}{3}&0\\
0&0&0&\alpha-1&0&0\\
\frac{\alpha}{3}&0&\frac{\alpha}{3}-1&0&\frac{\alpha}{3}&0\\
0&0&0&0&0&\alpha-1\\
\frac{\alpha}{3}&0&\frac{\alpha}{3}&0&\frac{\alpha}{3}-1&0\\
\end{bmatrix}.
\]
Then $\alpha_{\pm}=4/3{\pm}2\sqrt{2}/3i$ and 
\[
Spec({}^T\!{\bf W}_+)
=\Bigg{\{}0,0,\dfrac{4}{3}+\dfrac{2\sqrt{2}}{3}i,
-\dfrac{4}{3}-\dfrac{2\sqrt{2}}{3}i\Bigg{\}},\;
Spec({}^T\!{\bf W}_-)
=\Bigg{\{}0,0,\dfrac{4}{3}-\dfrac{2\sqrt{2}}{3}i,
-\dfrac{4}{3}+\dfrac{2\sqrt{2}}{3}i\Bigg{\}}.
\]
and 
\begin{equation*}
\begin{split}
Spec({\bf U}_+)&=\Bigg{\{}
{\pm}\Big{(}\dfrac{1}{\sqrt{3}}-\dfrac{\sqrt{2}}{\sqrt{3}}i\Big{)},
{\pm}\Big{(}\dfrac{1}{\sqrt{3}}-\dfrac{\sqrt{2}}{\sqrt{3}}i\Big{)},
1,\dfrac{1}{3}+\dfrac{2\sqrt{2}}{3}i,
-\dfrac{1}{3}-\dfrac{2\sqrt{2}}{3}i,-1 \Bigg{\}}-\{1,-1\},\\
Spec({\bf U}_-)&=\Bigg{\{}
{\pm}\Big{(}\dfrac{1}{\sqrt{3}}+\dfrac{\sqrt{2}}{\sqrt{3}}i\Big{)},
{\pm}\Big{(}\dfrac{1}{\sqrt{3}}+\dfrac{\sqrt{2}}{\sqrt{3}}i\Big{)},
1,\dfrac{1}{3}-\dfrac{2\sqrt{2}}{3}i,
-\dfrac{1}{3}+\dfrac{2\sqrt{2}}{3}i,-1 \Bigg{\}}-\{1,-1\}.
\end{split}
\end{equation*}
Thus we obtain: 
\begin{equation*}
\begin{split}
\sigma_r({\bf U})&=
\Bigg{(}\dfrac{1}{\sqrt{3}}-\dfrac{\sqrt{2}}{\sqrt{3}}i\Bigg{)}^{\!\!\HM^*}
{\cup}\Bigg{(}-\dfrac{1}{\sqrt{3}}+\dfrac{\sqrt{2}}{\sqrt{3}}i\Bigg{)}^{\!\!\HM^*}
{\cup}\Bigg{(}\dfrac{1}{3}+\dfrac{2\sqrt{2}}{3}i\Bigg{)}^{\!\!\HM^*}
{\cup}\Bigg{(}-\dfrac{1}{3}-\dfrac{2\sqrt{2}}{3}i\Bigg{)}^{\!\!\HM^*}.
\end{split}
\end{equation*}

Next, we state spectra of the Grover matrix. 
Let $\alpha=2$ and ${\bf U} = {\bf U}^{\rm Gro} $. 
Then ${\bf U}^{\rm Gro} $ is given as follows:  
\[ 
{\bf U}^{\rm Gro}=\begin{bmatrix}
0 & 1 & 0 & 0 & 0& 0 \\
\frac{-1}{3} & 0 & \frac{2}{3} & 0 & \frac{2}{3} & 0 \\
0 & 0 & 0 & 1 & 0& 0 \\
\frac{2}{3} & 0 & \frac{-1}{3} & 0 & \frac{2}{3} & 0 \\
0 & 0 & 0 & 0 & 0& 1 \\
\frac{2}{3} & 0 & \frac{2}{3} & 0 & \frac{-1}{3} & 0 \\
\end{bmatrix}.
\]
Furthermore, the transition matrix ${\bf T} $ of the simple random walk of $S_4 $ is 
\[
{\bf T}=\begin{bmatrix}
0 & 0 & 0 & 1\\
0 & 0 & 0 & 1\\
0 & 0 & 0 & 1\\
\frac{1}{3} & \frac{1}{3} & \frac{1}{3} & 0 
\end{bmatrix}.
\] 
Then 
\[
Spec({\bf T})
=\Bigg{\{}0,0,-1,1\Bigg{\}}
\quad \text{and} \quad 
Spec({\bf U}^{\rm Gro})=\Bigg{\{} \pm i, \pm i, 1 , -1 \Bigg{\}}.
\]
Thus we obtain: 
\[
\sigma_r({\bf U}^{\rm Gro} )=\{1\} {\cup} \{ -1\} {\cup} i^{\HM^*}.
\]
\end{exm}

\par\noindent
{\bf Acknowledgment.} The first author was partially supported by the Grant-in-Aid for Scientific Research (C) of Japan Society for the Promotion of Science (Grant No.21540116). The third author is partially supported by the Grant-in-Aid for Scientific Research (C) of Japan Society for the Promotion of Science (Grant No.19540154). We are grateful to K. Tamano, S. Matsutani and Y. Ide for some valuable comments on this work. 
\
\par

\begin{small}
\bibliographystyle{jplain}

\end{small}

\end{document}